# Tuning the 4*f*-state occupancy of cerium in highly correlated CeSi/ Fe multi-layers: a study by x-ray absorption spectroscopy

M. Münzenberg[*]

*IV. Physikalisches Institut, Universität Göttingen, Friedrich-Hund-Platz 1,*

*37077 Göttingen, Germany*

W. Felsch

*I. Physikalisches Institut, Universität Göttingen, Friedrich-Hund-Platz 1,*

*37077 Göttingen, Germany*

P. Schaaf

*II. Physikalisches Institut, Universität Göttingen, Friedrich-Hund-Platz 1,*

*37077 Göttingen, Germany*

PACS numbers: 75.70.Cn, 78.70.Dm, 78.67.Pt, 73.21.ac, 71.27.+a

*Author to whom correspondence should be addressed. Electronic address: mmuenze@gwdg.de

**Abstract**

Spectra of x-ray absorption and magnetic circular dichroism were measured at $M_{4,5}$(3d) and $L_{2,3}$(2p) edges of Ce in multilayers [Ce$_{1-x}$Si$_x$/Fe]xn, with $x$ between 0.1 and 0.65. The study uncovers the highly correlated nature of this layered system. An α-phase like electronic configuration of Ce is observed, with ordered magnetic moments on the 4$f$ and 5$d$ electrons induced by the interaction with Fe. Increasing the Si content reduces the strength of the hybridization between the 4$f$ and conduction-band states which is reflected in a growing occupation and magnetic polarization of the 4$f$ states. Variations of the shape and intensity of the $L_{2,3}$-edge dichroism spectra, discussed in a simple phenomenological model, show the importance of the exchange interaction between the Ce-4$f$ and 5$d$ electrons, spin polarized by the interaction with Fe at the interfaces, for the electronic structure of Ce at high Si concentration and low temperature. A model of the band structure of rare-earth transition-metal compounds permits to argue that magnetic order on the Ce 4$f$ electrons in the multilayers is due to different mechanisms: to hybridization of the Ce-4$f$ with the Fe-3$d$ states at low Si concentration and to intra-atomic 4$f$-5$d$ exchange at high Si concentration. This is at variance with magnetic order in the intermetallics CeSi$_{2-\delta}$ and CeSi which results from interaction between the localized 4$f$ magnetic moments mediated by the Si-derived ($s,p$) conduction electrons, in competition with the Kondo effect.

**I. Introduction**

There is broad consensus that the unusual properties of Ce and its compounds originate essentially from the interplay of strong correlations between the Ce 4f electrons and hybridization between 4f- and conduction-electron states. Phenomena as intermediate valency, pointing to non-integer occupation of the 4f shell, or heavy-fermion behavior, characterized by an extremely large contribution of the electronic specific heat, are prominent observations on such systems. But perhaps one of the most intriguing phenomena is the γ-α phase transition in Ce metal [1]. It occurs under pressure at room temperature or, at ambient pressure, on cooling to low temperature. Complexity of the underlying mechanisms is reflected in the lattice-symmetry preserving collapse of the atomic volume by about 15%, the loss of the magnetic moment and a profound modification of the electronic structure. The transition is driven by a sensible increase of the hybridization between 4f- and conduction-electron states which leads to a delocalization of the 4f states in the α phase [2,3]. Formally, hybridization may be defined in terms of a configuration interaction. This means that the ground state of Ce is written as a linear combination of $\left|4f^0v^{n+1}\right\rangle$, $\left|4f^1v^n\right\rangle$, and $\left|4f^2v^{n-1}\right\rangle$ electronic states ($v$ stands for the valence electrons), i.e. in terms of differently occupied 4f states. In compounds Ce shows γ- or α-phase-like character, depending of the hybridization strength. In α-like compounds like CeFe$_2$ or CeCo$_5$ the Ce atoms carry an ordered magnetic moment [4], in contradistinction to elemental α-phase Ce metal [1].

There is an ongoing interest in the electronic structure of the intermetallic compounds CeSi$_2$ and CeSi in the context of intermediate-valency, Kondo-lattice and heavy-fermion phenomena. Most of the information comes from various high-energy spectroscopies with high resolution [5-8]. In these compounds weak hybridization occurs between the 4f and conduction-band states, preferentially with those of p symmetry derived from Si. The Ce $\left|4f^1v^n\right\rangle$ state dominates the ground-state electronic configuration, with a small admixture of the $\left|4f^0\nu^{n+1}\right\rangle$ state [7,8]. A delicate balance between the Kondo effect quenching the localized 4f magnetic moments and the Rudermann-Kittel-Kasuya-Yosida (RKKY) interaction between these moments and the conduction electrons allows for nonmagnetic or magnetically ordered ground states; the former is realized in CeSi$_2$ [9], the latter in CeSi [8,10]. In a simple picture [11], this competition is governed by a single parameter, the effective exchange interaction $J$ between the between the 4f and conduction electrons which enters the characteristic energy scales both for the Kondo and RKKY interactions. In turn, $J$ depends on the 4f conduction-band hybridization strength which varies sensitively with the interatomic distances. The fragile character of the weakly hybridized electronic ground-state configuration of Ce in CeSi$_2$ is apparent in the behavior of off-stoichiometric CeSi$_{2-\delta}$ which orders magnetically at $\delta > 0.16$ [12,13]. Experiments on single-phase amorphous films of Ce$_{1-x}$Si$_x$ alloys have shown that structural disorder stabilizes the $\gamma$-like electronic configuration of Ce, on a large scale of Si concentrations $0.2 \leq x \leq 0.9$ [14]. Apparently the loss of the lattice periodicity reduces the (already weak) 4f-p hybridization. It is interesting to note that amorphous Ce$_{.33}$Si$_{.67}$ films and the bulk crystalline counterpart CeSi$_2$ present essentially the same local environment around the Ce ions [15]. A study on bulk amorphous Ce$_{1-x}$Si$_x$ alloys revealed Kondo behavior at low temperature in a wide range of Si concentrations, i.e. the absence of magnetic order and an enhanced electronic specific heat, similar as in heavy-fermion systems [16]. In contrast, Ce shows an $\alpha$-like electronic configuration in amorphous Ce$_{1-x}$Fe$_x$ alloy films and a magnetically ordered ground state. In this system the (itinerant) Ce-4f states are strongly hybridized with the Fe-3d states [17]. For example, in amorphous Ce$_{.34}$Fe$_{.66}$ the strong hybridization of the corresponding compound CeFe$_2$ is preserved in spite of the structural disorder. Differences in the physical properties are related to the different local structure.

We have previously shown by x-ray absorption spectroscopy (XAS) [18] and studies of x-ray magnetic circular dichroism (XMCD) [18,19] and resonant x-ray magnetic scattering (RXMS) [20] that the α-phase-like electronic structure of Ce is stabilized in Ce/Fe multilayers on a considerable depth scale near the interface and, in this configuration, carries an ordered magnetic moment on its 4$f$ and 5$d$ states. Both, the stabilization of α-like Ce and its magnetic polarization are driven by interface effects, i.e. by in-plane elastic strain resulting from the large mismatch between the interatomic spacings at the interfaces of the Ce and Fe layers, and by strong interfacial electronic correlations between the Ce-4$f$ and 5$d$ states and the Fe-3$d$ states, [17-19,21,22], respectively. Subsequent experiments have shown that the absorption of hydrogen in the Ce sublayers reduces the 4$f$-state conduction-band hybridization and leads to a relocalization of the 4$f$ states; i.e. the γ-phase-like electronic configuration of Ce is stabilized in multilayers $CeH_{2-\delta}$/Fe, due to an expansion of the $CeH_{2-\delta}$ lattice. The result is a profound modification of the electronic and magnetic properties [23-26]. The magnetization of the hydrided layers, for example, turns perpendicular to the layer plane at low temperatures. The orientational transition involves the Ce 4$f$ moments at the interfaces interacting with the Fe 3$d$ states via the Ce 5$d$ states [23,24]. The XMCD spectra probing the 5$d$ states show very unusual temperature- and orientation-dependent line shapes [27].

Here, we report on an experimental study of the electronic and magnetic properties of multilayers combining amorphous $Ce_{1-x}Si_x$ with bcc Fe. It was conceived, with the very different properties of the Ce/Fe and $CeH_{2-\delta}$/Fe multilayers in mind, to explore the effect of a progressive transition of the Ce-4$f$ states from itinerancy in the α phase (strongly hybridized) to localization in the γ phase (weakly hybridized), simply controlled by introducing an increasing amount of the ($s,p$) element Si into the Ce sublayers in the Ce/Fe heterostructures. In fact, according to the experiments on thin $Ce_{1-x}Si_x$ films mentioned above, ~20% of Si would stabilize the γ-phase-like electronic structure of Ce [14]. The single phase nature of the amorphous $Ce_{1-x}Si_x$ sublayers makes the multilayers well suited for such study. We note that a gradual transition between the two Ce configurations by hydrogenation of Ce is not experimentally achievable: the primary solubility of H in Ce is very small, the phase $CeH_{2-\delta}$ is pre-ferred. We found that in the multilayer environment with Fe the Ce-4f states in Ce$_{1-x}$Si$_x$ are more strongly hybridized with the conduction-band states than in the bare alloy layers; even at Si concentrations as high as 65% the hybridization is still perceptible: Ce presents the *α*-phase-like configuration, even though the occupation of the 4*f* states is high. This is in contrast to the bare layers. It points a stabilization by in-plane elastic strain at the interfaces and to involvement of the Ce-5d and Fe-3d states in the hybridization process with the 4f states. As in the Si-free layered Ce/Fe system, the *α*-like Ce phase is magnetically ordered. The observation once again illustrates the sensitivity of the 4*f* configuration in Ce systems to surface and interface effects, caused by modifications of the nature and strength of their hybridization with the electronic environment [28]. To explore this behavior it is important to obtain information on the correlation between the Ce-5*d*, -4*f* and Fe-3*d* electrons in this layered system. This is the intention of the present study.

As in the previous work on the Ce-based multilayers [17-19] a study of XAS and XMCD at the *M*$_{4,5}$ and *L*$_{2,3}$ edges of Ce were systematically performed to explore the ground-state electronic properties of Ce in these multilayers, related to the 4*f* and 5*d* electrons. Particular attention was focused on the *L*$_{2,3}$-edge spectra that provide information on the 5*d* electron states. This is crucial to understand the magnetic properties of the multilayers since the Ce-5*d* electrons play a key role in the subtle exchange process between the Fe-3*d* and Ce-4*f* magnetic moments. In addition, Ce *L*-edge XAS is a well-suited technique to help to clarify the electronic ground state configuration of Ce because under the influence of the 4*f*-electron core-hole Coulomb interaction the signatures of the mainly $\left|\underline{2p}4f^0v^{n+2}\right\rangle$ and $\left|\underline{2p}4f^1v^{n+1}\right\rangle$ final states (underscore signifies a hole state) are well separated in the spectra.

The paper is organized as follows. In Sect. II we present details of the experiments performed, as well as basic characteristics of the multilayer samples. We discuss the results of $^{57}$Fe Mössbauer spectroscopy which supplements the XMCD data. In Sect. III we present and discuss the XAS (III.A) and XMCD (III.B) data. Sect. III.A shows that by varying the Si con-centration in the $Ce_{1-x}Si_x$ sublayers, the occupancy of the Ce-4$f$ states may be tailored. In Sect. III.B we discuss the mechanisms underlying the magnetic polarization of the Ce-4$f$ and 5$d$ states. We present a phenomenological model of the Ce-$L_{2,3}$-edge XMCD spectra which shows that, at high Si content, their shape and amplitude are strongly influenced by the 4$f$-5$d$ exchange interaction.

## II. Experimental details and sample characteristics

Multilayers composed of the structural periods [$Ce_{1-x}Si_x$($t$Å)/Fe(30Å)] and alloy films $Ce_{1-x}Si_x$ were grown by computer-controlled ion-beam sputtering in an ultra-high vacuum chamber (base pressure $p < 5 \times 10^{-10}$ mbar). Highly pure sputtering gas Ar (6N) and target metals Ce (3N) and Fe (4N8) were used. The Ce-Si target was composed of a Ce plate partly covered by a 0.5 mm thick Si wafer with a regular array of holes with diameters adapted to the desired concentration of the $Ce_{1-x}Si_x$ films. The actual composition of these films was determined after preparation by Rutherford backscattering spectrometry (RBS), with an uncertainty of 2%. Partial pressures of reactive gases (e. g. $O_2$, $N_2$, $H_2O$) were below $10^{-10}$ mbar during the deposition process. Typical growth rates were 0.5 - 1.0 Å/s. Deposition was performed at liquid nitrogen temperature to minimize diffusion. For the multilayers, Kapton (12 μm) or Mylar foil (1.5 μm) coated with a 40 Å thick Cr buffer layer were used as substrates for the x-ray absorption experiments, to permit measurements in transmission mode, both in the hard and soft x-ray regime, respectively. Multilayer samples for x-ray diffraction, magnetometry and Mössbauer spectroscopy, and individual $Ce_{1-x}Si_x$ alloy films for comparative absorption measurements with hard x-rays were deposited on Si (100) wafers; they were equally precoated with a Cr buffer to warrant identical growth conditions and properties of the sample layers. The total thickness of the samples was near 4000 Å for the absorption measurements with hard x-rays at the Ce $L$ edges. In the case of the soft x-ray experiments at the Ce $M_{4,5}$ edges it was important to keep the multilayers sufficiently thin; a total thickness of 150 Å was chosen for the Ce part of the samples in order to have about 50% of transmission. All samples were covered with a 80 Å thick protection layer of Cr.

Structural characterization of the multilayers was performed by x-ray diffraction in $\Theta/2\Theta$ reflection geometry at small and large angles. In the low-angle regime sharp superlattice Bragg peaks appear up to 6$^{th}$ order. Analysis of the data employing Parratt fits [29] yields a rms structural roughness of nominally about 1 monolayer at the interfaces, independently of the $Ce_{1-x}Si_x$ sublayer thickness (We recall that the Fe-layer thickness is always the same (30 Å)). This shows that the layered stacks are well defined with good periodicity and sharp composition profiles, in agreement with the previous observation on Ce/Fe multilayers [30]. The large-angle diffraction spectra show that the 30 Å thick Fe sublayers grow in the bcc structure without texture. Application of the Debye-Scherrer formula yields a crystallite size of up to 26-30 Å in the growth direction, which corresponds to the film thickness. The $Ce_{1-x}Si_x$ sublayers are amorphous with α-phase-like Ce: there is a broad diffraction maximum near the (111)-reflection expected for α-Ce. The single $Ce_{1-x}Si_x$ films grow equally in an amorphous structure; but in this case the position of the broad diffraction maximum indicates γ-phase-like Ce.

As previously shown for Ce/Fe system [21] information on the interfacial interaction in the $Ce_{1-x}Si_x$/Fe multilayers can be derived by $^{57}$Fe Mössbauer spectroscopy; it supplements the results of x-ray absorption presented below. Fig. 1 shows such spectra for three samples with different Si concentrations measured by conversion-electron detection at room temperature. They are magnetically split, according to the ferromagnetic nature of the layers. The asymmetry of the lines and the broad background feature in the middle of the spectra indicate the presence of nonequivalent Fe sites. Data analysis was performed as previously, using a model suited to the layered structure of the samples [21]: the spectra were decomposed into a sextet representing the normal bcc part in core of the Fe layers, and a probability distribution of magnetic hyperfine fields attributed to Fe affected by the interfaces, $p(B_{hf})$ [31]. These interfacial distributions $p(B_{hf})$ are compared in Fig.1 on the right. For each multilayer, the con-tribution of the interface component to the total intensity of the spectrum corresponds to a nominal Fe-layer thickness of 8-10 Å per interface, depending on the Si concentration. This corroborates the previous observation on Ce/Fe multilayers [21]. It signifies a sizeable magnetic interaction of Fe with Ce and/or Si on a length scale exceeding the local thickness fluctuations derived from the small-angle x-ray diffraction data and mirrors a modification of the electronic structure of Fe near the interfaces. The hyperfine-field distributions underlying the interface component, $p(B_{hf})$, present characteristic structures. As argued before [21], these structures may be assigned to different parts of the interface. The region at low fields ($B_{hf}$ < 150 kG) represents Fe atoms next to the interface. For the multilayer with x = 0.1 there is a pronounced maximum at $B_{hf}$ = 45 kG. Here, the low-field part of $p(B_{hf})$ corresponds to 2.5 Å per interface, i.e. to nominally one monolayer of Fe. This value is compatible with the structural rms roughness. As previously for the Ce/Fe multilayers, we attribute this feature to a weakly magnetic 'interface alloy' [21]. The low-$B_{hf}$ signature in $p(B_{hf})$ becomes increasingly broader for a growing Si content. It suggests an increasing influence of Si $p$ states on the interfacial interaction in the multilayers. This is a point to which we shall return in connection with the Ce XMCD spectra discussed below. We note that for the highest Si concentration, $x$ = 0.65, a symmetric doublet has to be added to get a close fit of the Mössbauer spectrum. It indicates the presence of a small amount of paramagnetic Fe in this multilayer (1.5 Å per interface) and may point to the formation of a nonmagnetic silicide. The Mössbauer spectra, as well as the x-ray diffraction data, exclude the formation of Fe silicides (which usually form very easily) on a large depth scale in the multilayers. This must be due to the presence of Ce. The average magnetic hyperfine-field at the Fe sites in the $Ce_{1-x}Si_x$/Fe multilayers resulting from the total Mössbauer spectra is reduced by ~22% with respect to the bulk $\alpha$-Fe value (334 kOe), which confirms the previous result on the Ce/Fe multilayers [21].

The x-ray absorption spectra of the multilayers were measured in transmission mode, both on the energy-dispersive spectrometer D 11 of the DCI storage ring ($L(2p)$ edges of Ce) and on the SU 22 beam line on the asymmetric wiggler of the Super-ACO storage ring ($M_{4,5}$(3d) edges of Ce) at LURE (Orsay, France). Ce-$L_3$-edge absorption of single $Ce_{1-x}Si_x$ alloy films was measured in the total electron-yield mode. The XMCD spectra were recorded at constant beam helicity by varying the photon energy across a given absorption edge (in a single shot (D11) or step by step (SU22)) in magnetic fields up to 20 kOe oriented parallel and opposite to the x-ray propagation direction; they were applied at grazing incidence (30°, D11) or under an angle of 60° (SU 22) with respect to the layer planes and were high enough to saturate the magnetization of the samples. At SU 22 the field was inverted at each energy point. The circular polarization of the light is estimated to be 70% for the measurements at the $L$ edges, 25% at the $M_5$ and 17% at the $M_4$ edge. If $\bar{\mu}(H_{+,-})$ represents the normalized absorption coefficients for right-circularly polarized photons (helicity $-\hbar$) and for the two magnetic field directions $H_+$ parallel and $H_-$ antiparallel to the propagation direction of the light, the XMCD signal is defined as $\Delta\mu = \bar{\mu}(H_+)-\bar{\mu}(H_-)$, according to the generally accepted convention. All XMCD spectra presented here were scaled to 100% circular polarization. They represent an average over the $Ce_{1-x}Si_x$ sublayers. Limited sublayer thicknesses were investigated. It was outside the scope of this study to explore the depth of the Fe-induced magnetic polarization in the $Ce_{1-x}Si_x$ sublayers near the interfaces. Details of the data collection and the experimental setup are described elsewhere [32,33].

**III. Results and discussion**

**A. Isotropic x-ray absorption spectroscopy. Variation of the Ce 4$f$-state occupancy**

The 4f electronic configuration of Ce in bulk compounds has been extensively investigated by XAS at the $2p \rightarrow 5d$ ($L_{2,3}$) and $3d \rightarrow 4f$ ($M_{4,5}$) excitation thresholds. Recent examples concerning the intermetallic $CeSi_2$ can be found in references 6 and 7. Fig. 2 shows the Ce-$L_{1,2,3}$ and $M_{4,5}$-edge spectra of multilayers $[Ce_{1-x}Si_x/Fe]xn$ with different Si concentrations measured at room temperature. In each of the $L_2$ and $L_3$ as well as the $M_4$ and $M_5$ spectra two separate peaks, ~10 eV apart, are observed that reflect the nature of the 4*f* configuration as a result of final-state effects. This spectral signature shows that the Ce-4*f* configuration is α-phase like [17]. The gradual intensity reduction of the peak at higher photon energy as Ce is increasingly substituted by Si implies a modification of the 4*f* configuration. The 4*f*-related final-state effects lead to a two-step profile in the $L_1$ spectrum (Fig. 2) which involves a transition into the unoccupied Ce 6*p*-derived valence states. Much effort has been devoted to study the final-state features in such spectra with the aim of determining the ground-state properties. One of the central parameters in the description of this class of compounds is the effective 4*f*-state occupancy, $n_{4f}$. Since the $M_{4,5}$-edge spectroscopy is based on the transition of an electron into a 4*f* level it probes the 4*f* configuration, hence $n_{4f}$, in the ground state obviously more directly than the $L_{2,3}$-edge spectroscopy, in which it is involved only indirectly as a result of a final state effect due to the 4f-electron 2p-core-hole interaction [34,35]. But traditionally, this latter spectroscopy has been more frequently applied, and here we first focus on the study of the $L_2$ and $L_3$ edges.

### *1. $L_{2,3}$-edge spectra of Ce*

The presence of a core hole in the XAS final state makes it difficult to quantitatively extract the ground-state 4f occupation number, $n_{4f}$, directly from the $L_{2,3}$-edge XAS data [35,36]. To get an estimate of this number for the $Ce_{1-x}Si_x$/Fe multilayers and of its variation with the Si concentration, we use a phenomenological approach [35] widely employed by experimentalists (see, e.g., Ref. 6). The two-peak structure of the $L_2$ and $L_3$-edge signatures (Fig. 2, bottom panel) is interpreted as a superposition of two 'white-line' resonances associated with the final states $\left|2\underline{p}4f^1v^{n+1}\right\rangle$ (main peak) and $\left|2\underline{p}4f^0v^{n+2}\right\rangle$ (peak at higher energy). These features are superposed to a step-like absorption increase associated with an excitation into continuum states. The 4*f* ground-state occupancy $n_{4f}$ then is simply obtained from the relative intensity of the two white lines; their width $2\Gamma$ correlates with the width of the Ce 5*d* band. For this purpose the spectra are deconvoluted by the superposition of 2 broadened Lorentzian functions and 2 arctan step functions. Kotani *et al.* [37] and Malterre [38] have provided theoretical arguments that the $L_{2,3}$ near-edge absorption spectra of Ce yield a reliable image of its ground state configuration. Hence the phenomenological approach to the $L_{2,3}$ spectra, even if it is oversimplified, should provide a satisfactory measure of the 4f occupancy $n_{4f}$ in the ground state, as long as the $\left|\underline{2p}4f^0v^{n+2}\right\rangle$ signature has sufficient spectral weight.

## *2. $M_{4,5}$-edge spectra of Ce*

The Ce 4f ground-state occupancy $n_{4f}$ extracted from the $L_2$-edge spectra and the linewidth $2\Gamma$, which is correlated with the width of the 5d band, are reported in Fig. 3 as a function of the Si concentration $x$. For $x \leq 0.1$, $n_{4f}$ adopts the minimum value that is reached by $\alpha$-Ce metal exposed to an external pressure [39]. Hence, Ce is in a phase with an $\alpha$-like electronic configuration. With raising $x$ there is a continuous increase of $n_{4f}$ up to ~0.9. This reflects a progressive reduction of the 4f-state conduction-band hybridization, i.e. an increasing degree of localization of the 4f states in the multilayers. The progressive decrease of the number of Ce neighbors apparently weakens the *f-d* hybridization in the alloy sublayers. At the highest Si content, $x = 0.65$, $n_{4f}$ approaches that of the $\gamma$-like Ce phase which is close to 1. This is corroborated by the evolution of the linewidth $2\Gamma$ which decreases from above 10 eV expected for $\alpha$-like Ce in these multilayers to 6 eV expected when the $\gamma$-like Ce phase is approached [17].

The conclusions and results derived from the Ce-$L_{2,3}$ absorption spectra are confirmed by the evolution of the Ce-$M_{4,5}$ spectra of the multilayers (Fig. 2, top panel). The two-peak structure of both the $M_4$ and $M_5$ resonances corresponds to the final states $\left|3d^94f^2v^n\right\rangle$ (main contribution) and $\left|3d^94f^1v^{n+1}\right\rangle$ (satellite at the high photon-energy end); they are directly related to the initial states $\left|3d^{10}4f^1v^n\right\rangle$ and $\left|3d^{10}4f^0v^{n+1}\right\rangle$. This spectral shape differs from the $M_{4,5}$ resonances of typical $\gamma$-phase-like Ce systems [40]. In that case they exhibit a fine structure arising from the exchange splitting of the $\left|3d^9 4f^2 v^n\right\rangle$ final state which is well described by an atomic multiplet calculation of this configuration [41]. This fine structure is smeared out here due to the substantial hybridization between the 4f and conduction-electron states, which confirms the $\alpha$-phase-like character of Ce [42]. It is worth mentioning that in the corresponding spectra of weakly hybridized bulk $CeSi_2$ this fine structure is still visible, in addition to a weak $4f^1$ final state satellite above the main peak [7]. The theoretical description of the $M_{4,5}$ absorption thresholds in highly correlated Ce systems appears to be much clearer than that of the $L_{2,3}$ edges [34,43]. Following Gunnarsson and Schönhammer [34], it is possible to adjust the experimental $n_{4f}$ values, extracted from the relative weight of the $4f^1$ and $4f^0$ channel intensity contributions to the $M_{4,5}$ spectra, in order to determine a reliable value of $n_{4f}$ in the ground state. For the case of strong hybridization, we get the estimation of $n_{4f}$ presented in Fig. 3 as a function of the Si concentration. The numbers are close to the ones resulting from the $L_{2,3}$-edge profiles, with a systematic shift to lower values [17].

### *3. Single $Ce_{1-x}Si_x$ alloy films*

The electronic ground-state configuration of Ce is different in single alloy films $Ce_{1-x}Si_x$. This is visible in the Ce $L_3$-edge absorption spectra presented in Fig. 4 which confirms previous observations [14]. In contrast to the multilayers essentially only a single white-line resonance is observed already at $x = 0.1$. This means that here Ce is in the weakly hybridized $\gamma$-phase-like configuration. Apparently the intercalation of $Ce_{1-x}Si_x$ between Fe in the multilayers stabilizes the $\alpha$-like Ce configuration to a considerably high Si concentration. As suggested for the pure Ce/Fe multilayers [18], this may be due to mismatch-related strain at the interfaces.

For the single alloy films in Fig. 4 the amplitude of the white line increases by about 35% as the Si concentration is raised to $x = 0.65$. This visualizes a change in the electronic structure, presumably a charge transfer from the Ce-5*d* band to the *s-p* orbitals of Si. It can be seen in Fig. 4 that with increasing Si concentration an oscillation appears in the spectra above the absorption edge beyond 5750 eV. This fine structure (NEXAFS) is present in the spectra of the multilayers as well and indicates a structural change in the Ce environment on the scale of the atomic distances as it occurs when the number of Ce neighbors is reduced. Amorphous films of composition $Ce_{.33}Si_{.67}$ present a limit in this context since in this case, as was previously shown, each Ce ion is surrounded by 12 Si nearest neighbors [16].

**B. X-ray absorption magnetic circular dichroism spectroscopy**

*1. $M_{4,5}$-edge XMCD spectra of Ce. Mechanisms for the Ce 4f-electron magnetic polarization*

An important result of the previous study on Ce/Fe multilayers is that the Ce atoms in their α-phase-like electron configuration carry ordered magnetic moments of 5*d* and 4*f* origin due to interfacial interaction with Fe [17-19]. The 5*d* moment extends on a depth scale of at least 15 Å into the Ce layers and is coupled antiparallel to the Fe moment [17,18]; the 4*f* moment, equally antialigned to that of Fe, resides more close to the interface [17]. XMCD experiments at the $M_{4,5}$ (3*d*) absorption edges of Ce give access to the magnetic polarization of the 4*f* states. At room temperature such XMCD signals of the $Ce_{1-x}Si_x$/Fe multilayers (if they exist) are very small (< 0.5% of the white line) and beyond the experimental resolution. But at 10 K magnetic dichroism, reflecting an ordered 4*f* moment, is clearly visible (Fig. 5a). The XMCD signals at *x* = 0 agree with that of the previous study on Ce/Fe for a 10 Å thick Ce sublayer [19]. The dichroic amplitudes at the $M_4$ and $M_5$ edges are plotted in Fig. 5b as a function of the Si concentration *x*. They increase monotonically with *x* by up to about 50%. This increase is remarkable, it indicates that in the more diluted Ce-alloy sublayers the 4*f* electrons are magnetically polarized more effectively. A connection to the enhanced 4*f*-state occupation $n_{4f}$ in these sublayers (Fig. 3), i.e. to a reduced 4*f*-conduction-band hybridization, is obvious.

Note that we do not apply the dichroism sum rules [44] to the XMCD spectra in Fig. 5a to deduce the orbital and spin magnetic moments of the Ce-4f electrons in the ground state, $<L_z^{4f}>$ and $<S_z^{4f}>$. The result might be questionable due to (i) the varying 4f conduction-band hybridization, hence 4f state occupancy $n_{4f}$, as the Si content changes, (ii) the mixing of the state with total angular momentum $J = 7/2$ to the Ce ground state, demonstrated for the Ce/Fe multilayers [19]. This means also that we do not calculated the ratio $<L_z^{4f}>/<S_z^{4f}>$; its deviation from the free ion value is regarded as a sign of the 4f conduction-band hybridization [45-47]. Here, additional problems arise from uncertainties in the value of the so-called magnetic dipole operator $<T_z^{4f}>$ [46]. It adds with $<S_z^{4f}>$ to yield $<S_z^{4f}>_{eff}$, the quantity accessible by the spin sum rule [44].

The growing ordered magnetic moment of the Ce 4f electrons with increasing 4f-state occupation $n_{4f}$ can be ascribed to a change of the polarization mechanism as the 4f-conduction-electron hybridization diminishes. The multilayers considered here belong to the class of systems combining rare earths (REs) with magnetic transition metals (TMs). The electronic and magnetic structure of bulk compounds of such elements may be theoretically described by a model of Brooks and Johansson [48]; according to general arguments the underlying mechanisms may be applied to interface magnetism of the corresponding multilayers [49]. What is relevant here is that *hybridization between the itinerant RE 5d states and the spin-split TM 3d states,* with different strengths for the spin-up and spin-down states [48], give rise to an ordered 5d spin moment in the paramagnetic RE layers near the interface on a sizeable depth scale oriented antiparallel to that of the TM 3d moment. Ordered Magnetism on the RE 4f electrons originates from interactions with their electronic environment according to mechanisms dissimilar in the case of localized and itinerant 4f states.

(i) *4f states localized* : In this case, realized in the majority of the RE elements, the 4f states hybridize little with the TM 3d states. *Intra-atomic exchange coupling between the 4f and magnetically polarized 5d electrons* gives rise to an ordered 4f magnetic moment; it is indirectly coupled to the TM 3d moment [48]. This case is realized in Ce systems with a γ-phase like electronic configuration. The multilayers $CeH_{2-\delta}$/Fe [23,24] mentioned in the introduction are an example. In the present multilayer system, this mechanism is the dominating source for 4f-related XMCD at high Si concentration ($x = 0.65$). In this limit of high 4f-state occupancy the magnetic polarization of the 4f electrons, as our results obtained for the $CeH_{2-\delta}$/Fe multilayers [17] suggest, extends over the same length scale as that of the 5d electrons.

(ii) *4f states delocalized:* This case realized in α-phase-like Ce systems is more complex, a specific example is the intermetallic $CeFe_2$ [47,50,51]: the itinerant 4f states participate in band formation. In addition to 5d-3d hybridization now *4f-3d hybridization becomes important*, it generates an induced 4f spin moment on Ce in opposite direction to the 3d moment of Fe [48]. As in the Ce/Fe multilayers [17], this mechanism is the main source of the 4f-related XMCD in Fig. 5 in the limit of low 4f-state occupancy $n_{4f}$. This means, as we have shown previously, that in this limit the 4f electrons are only polarized, now due to a different mechanism, if Ce is in direct contact with Fe, i.e. at the direct interface [17]. As $n_{4f}$ increases, together with the Si concentration, the 4f-3d hybridization gradually weakens and intra-atomic exchange coupling between the 4f and 5d electrons, the latter ones spin polarized by hybridization with the Fe-3d band states, becomes increasingly important.

## 2. $L_{2,3}$-edge XMCD spectra of Ce. Influence of the 4f moment on the dichroic shape

It is essential to note that in all of these (bulk and layered) RE-TM systems the RE 5d band states are a central issue to understand the magnetic properties. XMCD spectra at the RE $L_{2,3}$(2p) edges are an important tool for obtaining information on the 5d-electron magnetism which is difficult to acquire separately by other methods. Fig. 6 displays the $L_2$ and $L_3$ XMCD spectra of Ce in the $Ce_{1-x}Si_x$ multilayer sublayers for different Si concentrations at 300 and 10 K. We note that in contrast to the $M_{4,5}$ edges a dichroic signal is clearly resolved at both L edges already at 300 K. As a signature of the 4f configuration mixing, each spectrum consists of two contributions, as the isotropic absorption spectrum (Fig. 2), with roughly the same splitting (~10 eV) of the $4f^1$ and $4f^0$ channels. Except for the sample with the highest Si content, $x = 0.65$ at 10 K, the signals are negative at the $L_2$ and positive at the $L_3$ edge. These signs indicate antiparallel orientation of the Ce-$5d$ and Fe-$3d$ magnetic moments for the definition of the XMCD signals employed here (Sec. II) [18]. At the Ce $L_1$ edge (transition $2s \rightarrow 6p$) magnetic dichroism could not be detected despite of the possibility of a magnetic polarization transfer by hybridization with the spin-split $5d$ states. In the light of a naïve expectation, in a single particle description [52] the evolution of the dichroic $L_2$ and $L_3$ signals with Si concentration and temperature is largely anomalous. An exception is the case with the lowest Si concentration $x = 0.1$. Here the XMCD signals increase when the temperature is reduced from 300 to 10 K while the spectral shape is preserved. Furthermore, at both temperatures the ratio of the XMCD signal amplitudes, $|L_2/L_3|$, is close to the statistical value of 1 imposed by the degeneracy of the $2p_{1/2}$ and $2p_{3/2}$ core states, in a simple picture of the $2p \rightarrow 5d$ dipole transition, with negligible spin-orbit coupling in the $5d$ final state [53]. This behavior agrees with the previous observations on the Ce/Fe multilayers; it shows that the orbital contribution to the $5d$ magnetic moment is almost zero, i.e. the moment is essentially of pure spin origin [18]. It can be seen in Fig. 6 that the situation is different for the other multilayers more rich in Si, $x > 0.1$. At 300 K, the amplitudes of the dichroic signals at both the $L_2$ and $L_3$ edges decrease monotonously with growing Si concentration, while the spectral shape does not change. This decrease is intriguing in view of the increasing amplitudes of the $M_4$- and $M_5$-edge XMCD signals (Fig. 5b), i.e. a growing magnetic polarization of the $4f$ electrons, and, as we shall argue below, a progressive strengthening of the $4f$-$5d$ exchange interaction. Further irregularities in the XMCD spectra appear when the temperature is reduced to 10 K. For a given Si concentration $x > 0.1$, the amplitudes do not simply grow. At the $L_2$ edge (Fig. 6 a), the spectral shape changes, most dramatically for $x = 0.65$ where it is derivative like owing to the reversed sign of the low-energy peak related to the $4f^1$ channel. In contrast, the shape of the $L_3$ edge spectra (Fig. 6 b) is essentially preserved.

The complex XMCD spectrum at the Ce-$L_2$ edge of the multilayer Ce$_{.35}$Si$_{.65}$/Fe at 10 K is similar to the Ce $L_2$-edge dichroic spectra of the multilayers CeH$_{2-\delta}$/Fe [27] which have a $\gamma$-phase-like Ce-$4f$ configuration. This suggests that the increased localization of the Ce-$4f$ states in the Si-rich Ce$_{.35}$Si$_{.65}$ sublayer plays a role in the anomaly. Let us note that in the hydride multilayer system shape and amplitude of the XMCD signals vary strongly with the angle between the magnetic field, which is parallel to the beam and the layer normal. The effect is related to the magnetocrystalline anisotropy of the multilayers which induces a perpendicular magnetization orientation at low temperatures [24]. No anisotropy in the Ce $L$-edge XMCD spectra in any of the Ce$_{1-x}$Si$_x$/Fe multilayers could be detected [54], in agreement with the observation that its global magnetization lies in the layer plane at all temperatures.

In spite of their importance it is not straight forward to draw instructive information from the experimental XMCD spectra at the RE $L_{2,3}(2p)$ edges [55]. For example, they do not simply provide information on the spin-resolved $5d$ density of states; even the sign of XMCD intensities could not be predicted by a basic theory. Complexities arise from the intra-atomic exchange interaction between $5d$ and $4f$ electrons which leads to a spin-dependent enhancement of the $2p$-$5d$ radial matrix element in the photoabsorption process [56]. It means that the $L$-edge XMCD depends on the $4f$ states. Model calculations considering the $5d$-$4f$ exchange interaction [57-59] have shown that it is indispensable for a proper interpretation of the spectra; it is differently effective at the $L_2$ and $L_3$ edges. The calculations were successful, for example, in describing basic trends in the signs and widely varying ratios of dichroic signals in RE-based intermetallics [55]. Nevertheless, in spite of these achievements and more recent theoretical progress [51] understanding of these spectra remains is limited to date.

Since a rigorous theoretical description of the RE $L_{2,3}$-edge dichroism does not exist we use a physically transparent phenomenological procedure to model the evolution of the Ce-$L_{2,3}$ XMCD spectra of the $Ce_{1-x}Si_x$/Fe multilayers with the Si content and temperature. The model is motivated by the existing calculations [58,59] and makes use of the method [35] applied above to analyze the isotropic XAS spectra at the $L_{2,3}$ edges of the RE (Sect. III.A). Previously invoked to analyze the $L_{2,3}$ XMCD spectra of Ce in the $CeH_{2-\delta}$/Fe multilayers, it has lead to basic insights concerning the spectral shape and intensity [27]. We assume that the magnetic interaction of the RE 5$d$ states with Fe at the interfaces in the multilayers has two effects that control the $L_{2,3}$ XMCD spectra of the RE; they are represented by 2 parameters, $\alpha$ and $\beta$. (The model is only briefly addressed here. For a more detailed discussion, see Ref. 27).

(i) The interfacial interaction results in a difference in the spectral weight for unoccupied spin-up and spin-down 5d states that may be ascribed to *an 'effective magnetic polarization'*

$$\alpha = (\Delta\rho/\rho + \Delta M/M). \qquad (1)$$

It determines the dichroic intensity corresponding to the $2p \rightarrow 5d$ electric dipole ($E_1$) transition, in an extension of an independent particle description [60,61], by

$$\Delta\mu = P_e \alpha, \qquad (2)$$

where $\Delta\mu$ is normalized to the isotropic absorption coefficient (Sect. II). The relation combines the difference of unoccupied spin-up (majority) and spin-down (minority) 5$d$ states per atom in the ground state, $\Delta\rho = \rho^{\uparrow} - \rho^{\downarrow}$, here induced by 5$d$-3$d$ hybridization, and the difference between the radial parts of the matrix element for the transition to these states, $\Delta M = M^{\uparrow} - M^{\downarrow}$; $\Delta M$ results from the exchange interaction between the excited 5$d$ electron and the 4$f$ electron, $\varepsilon_{4f\text{-}5d}$, which generates a spin-dependent contraction of the radial part of the 5$d$ wave function that is proportional to $\varepsilon_{4f\text{-}5d}$ [62]. $P_e$ is the spin polarization of the photoelectron (-0.5 and +0.25 for the $L_2$ and $L_3$ edges, respectively). For a net 5$d$ magnetic moment oriented parallel to the x-ray propagation direction, we have $\Delta\rho<0$ and $\Delta M>0$. ($\Delta\rho$ is in the range of a few percent at the RE-$L_{2,3}$ edges whereas $\Delta M$ can rise to 20-30% [61].) Hence, due to the different sign of $\Delta\rho$ and $\Delta M$, an enhancement of $M^\uparrow$ may reverse the sign of the XMCD signal (Eq. 2). Eq. 1 mixes band structure effects ($\Delta\rho$) and intraatomic final-state interaction ($\Delta M$). In spite of its intra-atomic character it should be a good approximation for metallic systems due to the local nature of the $L_{2,3}$ XMCD.

(ii) *There is a splitting in energy for the final states of the excited spin-up and spin-down 5d electron, we denote by $\beta$.* One part of the effect is due to 4f-5d exchange interaction $\varepsilon_{4f-5d}$ [58,59] as in the case of the parameter $\alpha$, another part to the exchange interaction between the RE-5$d$/Fe-3$d$ hybridized final state and the 2$p$ core hole [63]. Note that the parameters $\alpha$ and $\beta$ are correlated due to their dependence on $\varepsilon_{4f-5d}$.

The two parameters $\alpha$ and $\beta$ are used to model the $L_{2,3}$ XMCD intensity as a function of energy $E$ by following the approach of Röhler [35] for the deconvolution of the x-ray absorption spectra at the RE $L_{2,3}$ edges (see Sect. III.A). The white-line resonances for both spin channels in the $E_1$ transition to the spin-up and spin-down unoccupied 5$d$ states are approximated by Lorentzian functions with an amplitude $A_L$ and a width $2\Gamma$; their difference is used to represent the XMCD intensities at the $L_3$ edge (upper sign) and $L_2$ edge (lower sign):

$$\Delta\mu(E, A_L, 2\Gamma) = \pm L(E, A_L, 2\Gamma) \mp L(E-\beta, A_L(1-\alpha), 2\Gamma+\beta) \quad (3)$$

The different signs express the opposite orientation of the spin of the photo-excited electron at the $L_3$ and $L_2$ edges. For a general discussion of the effect of the two parameters $\alpha$ and $\beta$ in Eq. (3) see Ref. 27.

There is little doubt that the anomalies of the XMCD spectra at the $L_{2,3}$ edges of Ce addressed above (Fig. 6), like the decrease in intensity with increasing Si content $x$ or the change in shape with decreasing temperature (at $x > 0.1$) can be attributed to the 5$d$-4$f$ exchange interaction. In fact, these effects are accompanied by a growing magnetic polarization of the 4$f$ electrons (Fig. 5) which is related to an increasing 4f-state occupancy $n_{4f}$ (Fig. 3). We have analyzed the dichroic Ce $L_{2,3}$-edge spectra of the $Ce_{1-x}Si_x$/Fe multilayers by the phenomenological model (Eq. 3). As an example, Fig. 7 shows the results for the $L_2$ edge, for 2 Si concentrations, $x = 0.4$ and $0.65$, at various temperatures. Separate fits are applied for the XMCD signatures of the $4f^1$ channel at the absorption edge, using Eq. 2, and of the $4f^0$ channel at ~10 eV above the edge, using an additional Lorentzian function [35]. The results of the analysis of the white-line spectra were used as initial parameters (width $2\Gamma$, amplitudes $A_L$ and the ratio of the $4f^1$ and $4f^0$ contributions); thus only $\alpha$ and $\beta$ and the polarization of the $4f^0$ contribution are used as free parameters. Inspection of Fig. 7 reveals that the evolution of the dichroic spectra with temperature lowered from 300 to 10 K is well described. In fact, the negative signal at the at the Ce-$L_2$ edge results from Eq. 1 if $|\Delta M/M| < |\Delta\rho/\rho|$ and the projected Ce-5d magnetic moment is oriented opposite to the x-ray propagation direction, which is parallel to the external magnetic field. Since in the experiment the Fe magnetic moment is aligned along the field this antiparallel moment orientation is an expectation of the band-structure theory of RE-TM compounds [48] addressed above. Note the decrease in amplitude of the $4f^1$ signal for $x = 0.4$ (left panel), and the gradual variation of this feature from predominantly negative to strongly positive XMCD for $x = 0.65$ (right panel). The values of the parameters $\alpha$ and $\beta$ extracted from the analysis are presented in Fig. 8. The vanishing $\beta$ for $x = 0.1$ in the entire temperature range suggests that for this Si concentration the 4f-5d exchange interaction is negligible. This is related to the important 4f valence-band hybridization in this case and means that the XMCD signal represents the spin polarization of the Ce-5d band in the ground state, i.e. $\alpha \approx (\Delta\rho/\rho)$. For $x = 0.4$, a slight decrease of $\alpha$ at 10 K goes along with a finite $\beta$. It means that the 4f-5d exchange interaction is effective, hence the spin dependence of the matrix element for the 2p→5d transition, $\Delta M$, that competes with the 5d spin polarization $\Delta\rho$ gains in importance; $\alpha$ is reduced (Eq. 1). The reduced amplitude of the $4f^1$-related signal in the XMCD spectrum (Fig. 7) is a consequence. For $x = 0.65$, 4f-5d exchange is even more important. As an effect of a substantial $\Delta M$ contribution, $\alpha$ is very small at 300 K and changes sign near 50 K. As $\alpha$ crosses zero, i.e. for $|\Delta M/M| < |\Delta\rho/\rho|$, the $4f^1$-related XMCD signal becomes derivative-like, emphasizing the role of $\beta$ growing toward low temperature. To conclude, the variation of the parameters $\alpha$ and $\beta$ of our model with temperature and Si content (Fig. 8) extracted from a comparison with the XMCD spectra in Fig. 7 clearly show the importance of the 4f-5d exchange interaction for the Ce-$L_2$ spectra of the Ce$_{1-x}$Si$_x$/Fe multilayers for high Si concentration and low temperature. It goes along with a high 4f-state occupancy $n_{4f}$ (Fig. 3), i.e. an increased localization of the Ce-4f states. In that sense the similarities of the dichroic spectra of the multilayer Ce$_{.35}$Si$_{.65}$/Fe (Fig. 7) and the hydrided multilayers CeH$_{2-\delta}$/Fe where the Ce electron configuration is $\gamma$-phase like [27] are not surprising. The experiments reveal that the strength of the 4f-5d exchange interaction can be controlled by varying the composition of the Ce$_{1-x}$Si$_x$ sublayers. Along with the modified exchange interaction, the mechanism of the magnetic 4f-electron polarization changes: from 4f-3d hybridization at low Si concentration $x$ to intra-atomic 4f-5d exchange at high Si concentration. In both cases, interaction with Fe is the source of ordered 4f magnetism on Ce in the multilayers. This is at variance with magnetic order in the intermetallics CeSi$_{2-\delta}$ and CeSi which results from interaction between the localized 4f magnetic moments mediated by the ($s,p$) conduction electrons via the RKKY mechanism [8,12].

Inspection of Fig. 6b shows that the XMCD spectra at the $L_3$ edge of Ce are affected by the 4f-5d exchange interaction, too: their intensity, for example, decreases with increasing Si content. However, the spectral shape is not modified at low temperature. In particular, there is no sign reversal as in the $L_2$-edge spectrum at x = 0.65 Si (Fig. 7). It indicates that the 4f-5d exchange is less effective at this absorption threshold, in agreement with our observation on the CeH$_{2-\delta}$/Fe multilayers [27] and theoretical prediction for the enhancement of the relevant dipole matrix element $M$ [51]. The spectra can be fitted to our phenomenological model. This result is not presented here, since not much new information can be learned from this analysis.

**IV. Conclusion**

The present study discloses the highly correlated nature of $Ce_{1-x}Si_x$ layers in a multilayer environment with Fe. Their properties are very different from those of the compounds $CeSi_{2-\delta}$ and CeSi. We have demonstrated that the configuration of the 4*f* states of Ce in the layered structure may be tuned from delocalized at low Si content *x* to more localized at high *x*. This is visible in an increase of the 4f-state occupancy. An important parameter is the hybridization of the 4*f*- and conduction-band states. It changes from weak strength in the bare $Ce_{1-x}Si_x$ layers where the Si-derived 3*p* states are implicated in the mixing process to higher strength in the multilayers where the Ce-5*d* and Fe-3*d* states are involved. Furthermore, it decreases in the multilayers with increasing Si concentration. The reduced 4*f* conduction-band hybridization favors the intra-atomic exchange interaction of the 4*f* electrons with the 5*d* electrons of Ce, spin-split by the interaction with Fe. This is the source of ordered 4*f* magnetism of Ce in the $Ce_{1-X}Si_X$ sublayers at high Si concentration. In contrast, hybridization of the Ce-4*f* states with the Fe-3*d* states at the interface induces magnetic order on Ce at low Si concentration.

These results highlight the unique possibility of XAS and XMCD spectroscopy, here in particular on the $L_{2,3}$ absorption edges of Ce, to explore the electronic and magnetic properties of this element in the multilayers. It is shown that a simple phenomenological model permits to parameterize the $L_{2,3}$-edge XMCD spectra for which a rigorous theory awaits to be established. Trends in the variation of shape and intensity of the spectra with Si concentration and temperature can be traced down then to the 4*f*-5*d* exchange interaction which is well known to have a strong bearing on the *L*-edge XMCD of RE systems [55,58,59].

**Acknowledgements**

Special thanks are due to G. Krill, J.P. Kappler, E. Dartyge, and F. Baudelet in Orsay for stimulative discussions and support at the beamline. The work was performed under the auspices of the Deutsche Forschungsgemeinschaft within SFB 345.

**Figure captions**

Fig. 1. (Color online) Left: $^{57}$Fe Mössbauer conversion electron spectra at room temperature of multilayers Ce$_{1-x}$Si$_x$/Fe with different Si concentrations $x$. Solid lines: fits described in the text. Right: magnetic hyperfine-field distribution $p(B_{hf})$ corresponding to the contribution of the interfaces to the spectra.

Fig. 2. (Color online) X-ray absorption spectra at the $M_{4,5}$ (top) and $L_{1,2,3}$ (bottom) edges of Ce in multilayers Ce$_{1-x}$Si$_x$/Fe with different Si concentrations $x$. The $M$-edge spectra are normalized to the amplitude of the M$_5$-edge spectra arbitrarily set to unity. For the $L$-edge spectra, the edge jumps are normalized to 0.5 at the $L_3$, and to 0.25 at the $L_2$ and $L_1$ edges. The decomposition of the spectra is demonstrated by the superposition of 2 Lorentzians ($M_5$ edge, top, blue (doted) lines)) and of 2 arctan functions plus 2 Lorentzians ($L_3$ edge, bottom, blue (dotted) lines). The $4f^1$ and $4f^0$ initial channels are indicated for the $M_5$ and $L_3$ edges by the vertical dashed lines.

Fig. 3. Occupation of the Ce $4f$ states, $n_{4f}$, extracted form the analysis of the $L_2$ and $M_5$ absorption edges (solid circles and squares, respectively) for multilayers Ce$_{1-x}$Si$_x$ (30Å)/Fe(30Å) as a function of the Si concentration $x$. Also shown: width of the 'white lines', $2\Gamma$, resulting from analysis of the $L_2$ edge (open circles). The value marked with brackets is for a Ce$_{1-x}$Si$_x$-sublayer thickness of 10 Å.

Fig. 4. X-ray absorption spectrum at the Ce-$L_3$ edge of a 500 Å thick Ce$_{1-x}$Si$_x$ films with different Si concentrations (total electron yield (TEY) detection) and of a multilayer Ce$_{0.9}$Si$_{0.1}$(30Å)/Fe(30Å) as a reference (bottom curve).

Fig. 5a. XMCD spectra at the Ce-$M_{4,5}$ edges of multilayers Ce$_{1-x}$Si$_x$/Fe with different Si concentrations $x$ measured at 10 K. The spectra are normalized to the isotropic $M_5$-edge spectra and 100% polarization rate; they represent an average over 4 spectra.

Fig. 5b. Peak amplitudes of the XMCD signals at the Ce-$M_4$ (negative peak) and $M_5$ edges of multilayers Ce$_{1-x}$Si$_x$/Fe as a function of the Si concentration $x$.

Fig. 6. Normalized XMCD spectra at the $L_2$ edge (top) and $L_3$ edge (bottom) of Ce for the multilayers Ce$_{1-x}$Si$_x$/Fe with different Si concentrations $x$ measured at 300 K (left) and 10 K (right).

Fig. 7. (Color online) Analysis of the $L_2$-edge XMCD spectra of Ce in two multilayers Ce$_{1-x}$Si$_x$/Fe with 40% (left) and 65% Si (right) at different temperatures using the model described in the text (Eqs. (1)-(3)). Shown are the fit for the first transition ($4f^1$ channel) at lower photon energy (black continuous line, yielding the parameters $α$ and $β$), the transition at higher energy ($4f^0$ channel, red (dotted) line), and their sum (blue (continuous) line).

Fig. 8. Parameters $α$ and $β$ as a function of temperature resulting from the fits of the phenomenological model (Eqs. (1)-(3)) to the Ce-$L_2$-edge XMCD spectra of multilayers Ce$_{1-x}$Si$_x$/Fe with different Si concentrations $x$. Solid lines: exponential functions serving as guides to the eye (see Ref. 54).

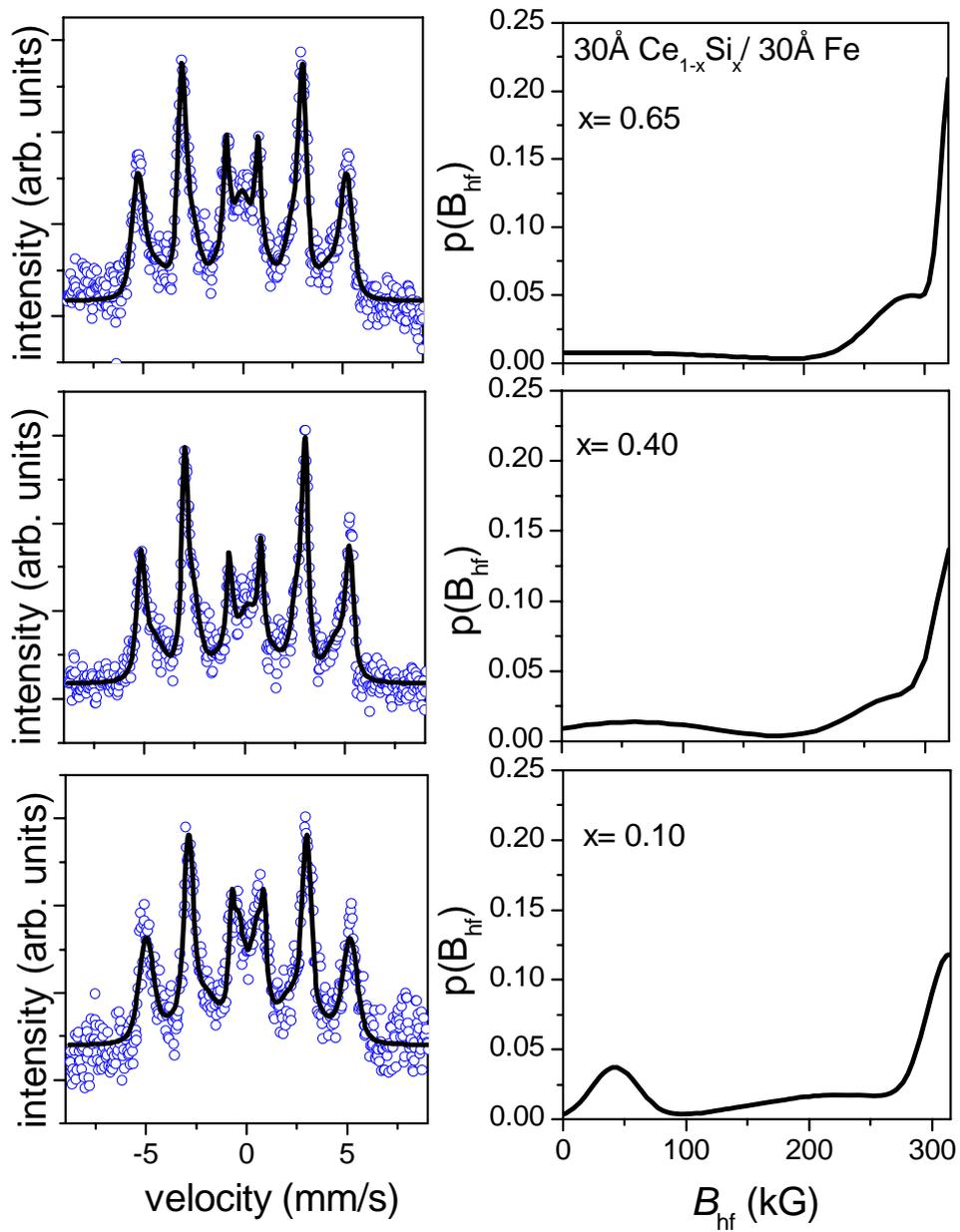

Fig. 1. (Color online) Left: $^{57}$Fe Mössbauer conversion electron spectra at room temperature of multilayers $Ce_{1-x}Si_x$/Fe with different Si concentrations $x$. Solid lines: fits described in the text. Right: magnetic hyperfine-field distribution $p(B_{hf})$ corresponding to the contribution of the interfaces to the spectra.

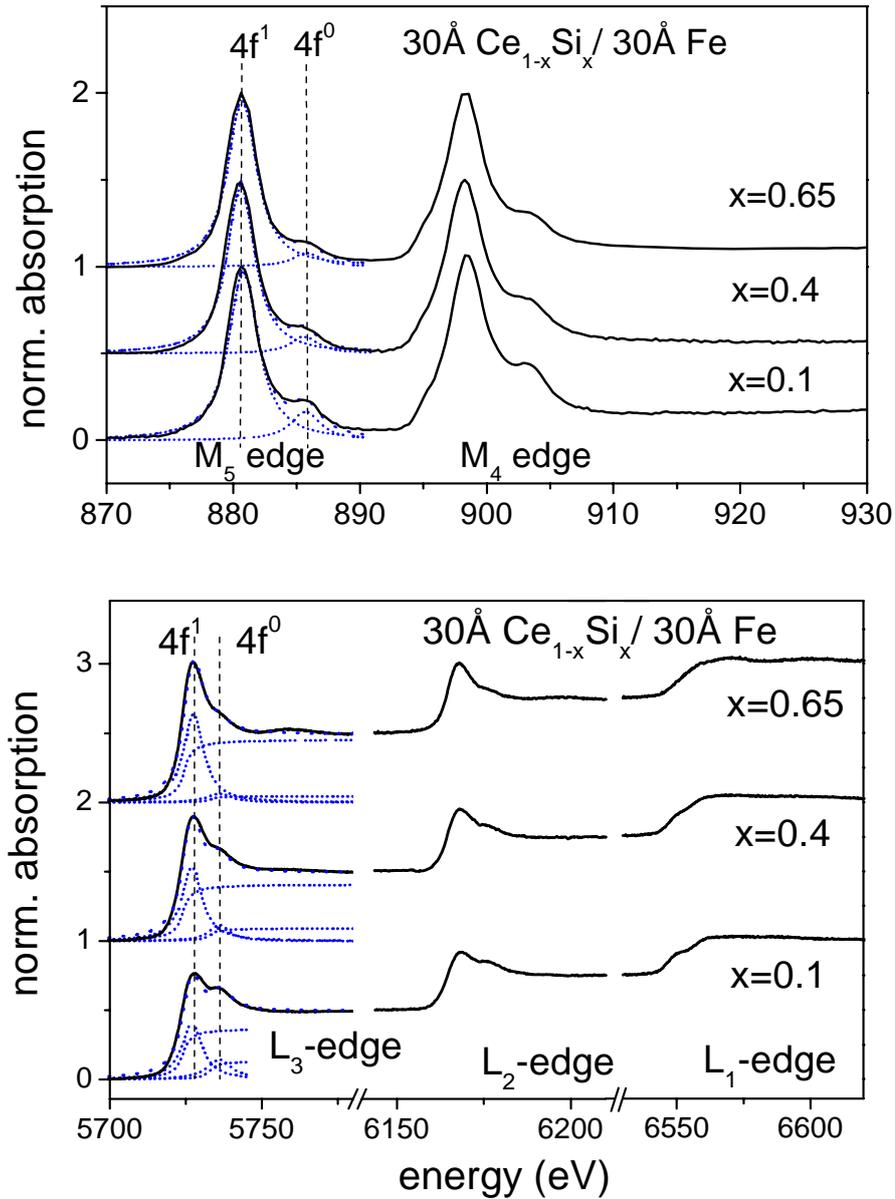

Fig. 2. (Color online) X-ray absorption spectra at the $M_{4,5}$ (top) and $L_{1,2,3}$ (bottom) edges of Ce in multilayers $Ce_{1-x}Si_x$/Fe with different Si concentrations $x$. The $M$-edge spectra are normalized to the amplitude of the $M_5$-edge spectra arbitrarily set to unity. For the $L$-edge spectra, the edge jumps are normalized to 0.5 at the $L_3$, and to 0.25 at the $L_2$ and $L_1$ edges. The decomposition of the spectra is demonstrated by the superposition of 2 Lorentzians ($M_5$ edge, top, blue (dotted) lines)) and of 2 arctan functions plus 2 Lorentzians ($L_3$ edge, bottom, blue (dotted) lines). The $4f^1$ and $4f^0$ initial channels are indicated for the $M_5$ and $L_3$ edges by the vertical dashed lines.

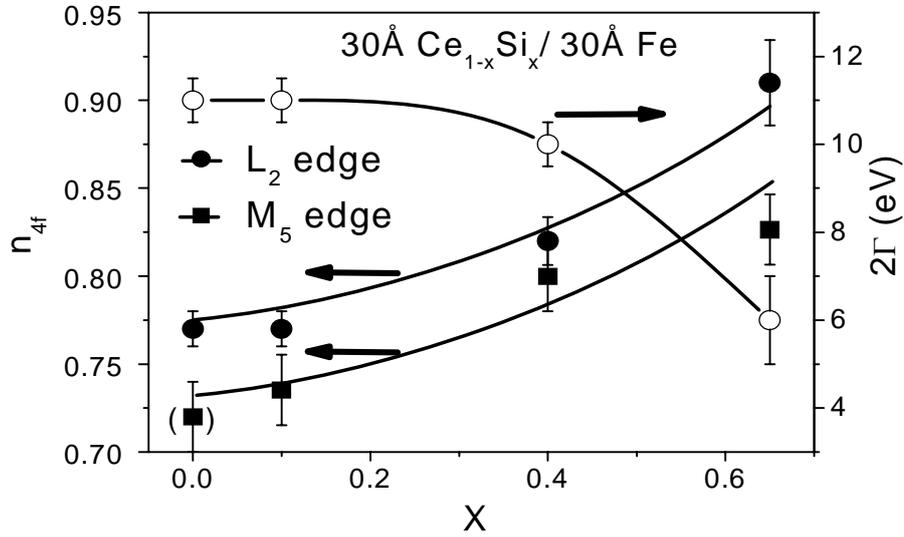

Fig. 3. Occupation of the Ce 4*f* states, $n_{4f}$, extracted form the analysis of the $L_2$ and $M_5$ absorption edges (solid circles and squares, respectively) for multilayers $Ce_{1-x}Si_x$ (30Å)/Fe(30Å) as a function of the Si concentration *x*. Also shown: width of the 'white lines', $2\Gamma$, resulting from analysis of the $L_2$ edge (open circles). The value marked with brackets is for a $Ce_{1-x}Si_x$-sublayer thickness of 10 Å.

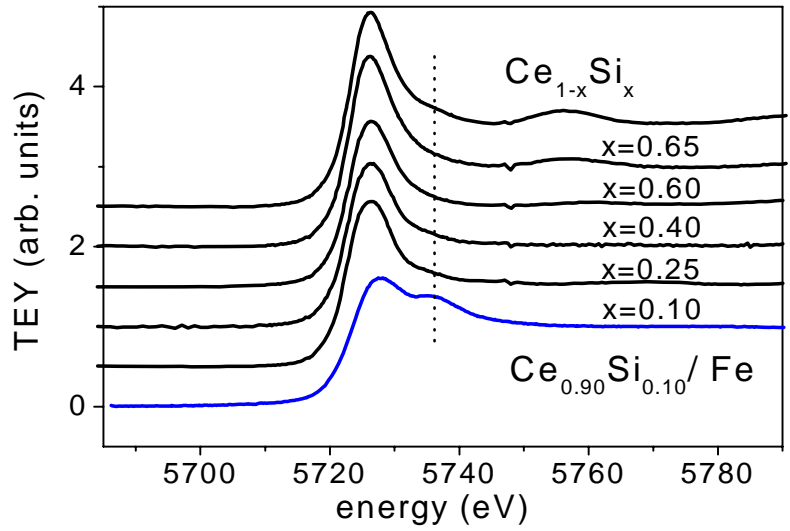

Fig. 4. X-ray absorption spectrum at the Ce-$L_3$ edge of a 500 Å thick $Ce_{1-x}Si_x$ films with different Si concentrations (total electron yield (TEY) detection) and of a multilayer $Ce_{0.9}Si_{0.1}$(30Å)/Fe(30Å) as a reference (bottom curve).

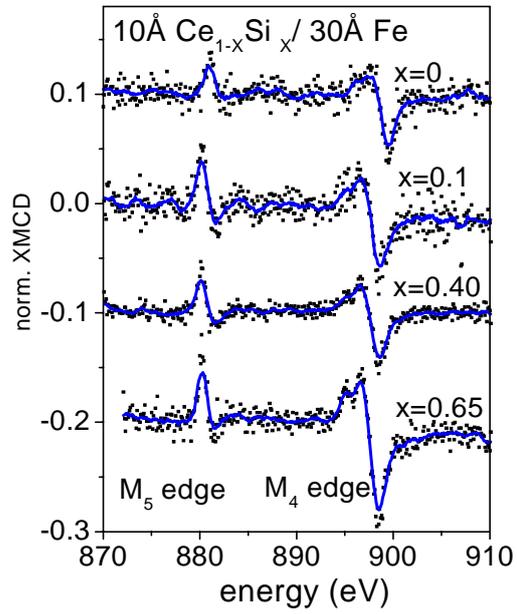

Fig. 5a. XMCD spectra at the Ce-$M_{4,5}$ edges of multilayers $Ce_{1-x}Si_x$/Fe with different Si concentrations $x$ measured at 10 K. The spectra are normalized to the isotropic $M_5$-edge spectra and 100% polarization rate; they represent an average over 4 spectra.

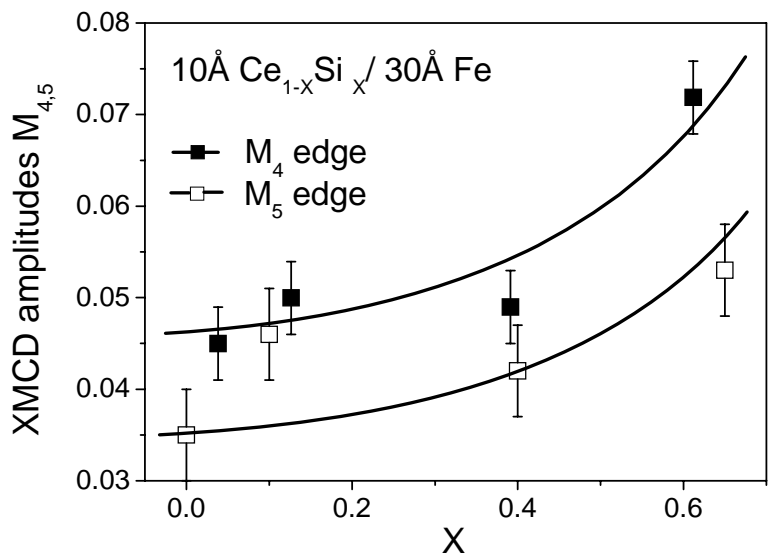

Fig. 5b. Peak amplitudes of the XMCD signals at the Ce-$M_4$ (negative peak) and $M_5$ edges of multilayers $Ce_{1-x}Si_x$/Fe as a function of the Si concentration $x$.

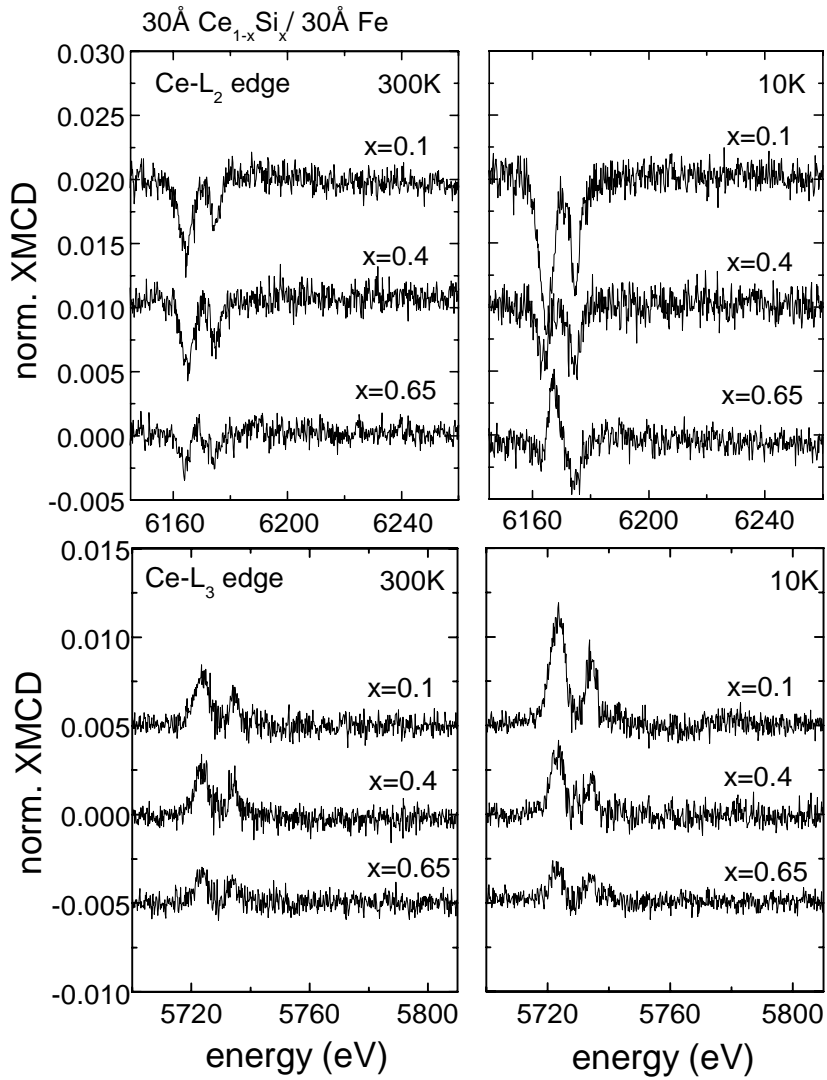

Fig. 6. Normalized XMCD spectra at the $L_2$ edge (top) and $L_3$ edge (bottom) of Ce for the multilayers Ce$_{1-x}$Si$_x$/Fe with different Si concentrations $x$ measured at 300 K (left) and 10 K (right).

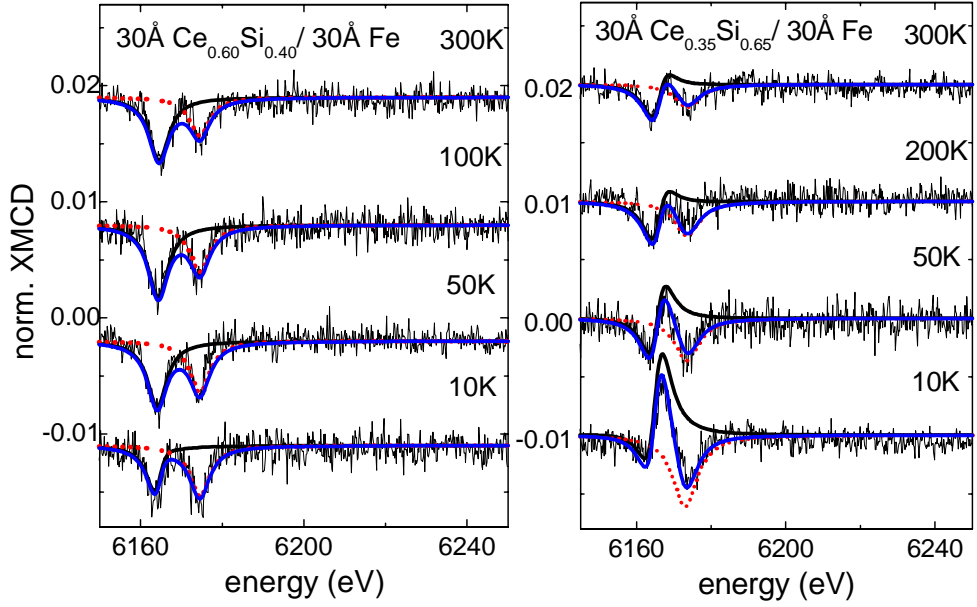

Fig. 7. (Color online) Analysis of the $L_2$-edge XMCD spectra of Ce in two multilayers Ce$_{1-x}$Si$_x$/Fe with 40% (left) and 65% Si (right) at different temperatures using the model described in the text (Eqs. (1)-(3)). Shown are the fit for the first transition ($4f^1$ channel) at lower photon energy (black continuous line, yielding the parameters $\alpha$ and $\beta$), the transition at higher energy ($4f^0$ channel, red (dotted) line), and their sum (blue (continuous) line).

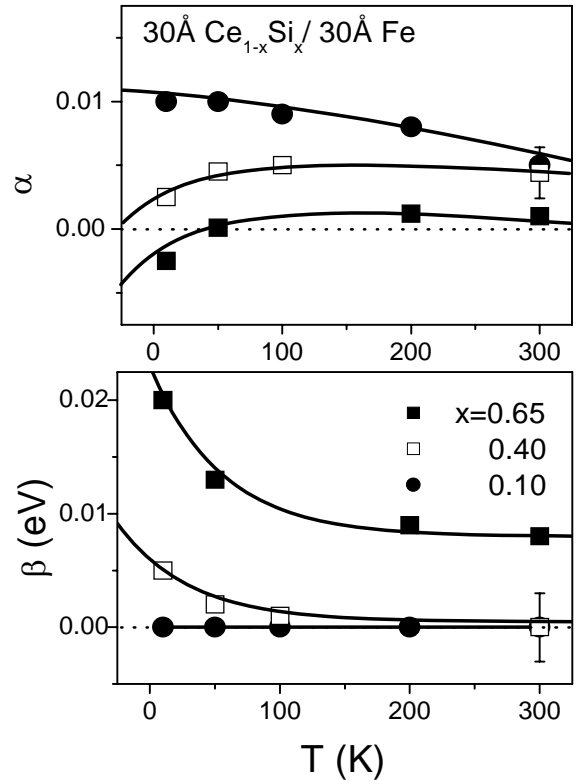

Fig. 8. Parameters $\alpha$ and $\beta$ as a function of temperature resulting from the fits of the phenomenological model (Eqs. (1)-(3)) to the Ce-$L_2$-edge XMCD spectra of multilayers Ce$_{1-x}$Si$_x$/Fe with different Si concentrations $x$. Solid lines: exponential functions serving as guides to the eye (see Ref. 54).